\documentstyle[epsfig]{aipproc}

\begin{document}
\title{From Superstrings to M Theory}

\author{John H. Schwarz$^*$}
\address{$^*$California Institute of Technology\thanks{Work
supported in part by the U.S. Dept. of Energy under Grant No.
DE-FG03-92-ER40701.}\\
Pasadena, California 91125}

\maketitle

\begin{abstract}
In this talk  I will survey some of the basic
facts about superstring theories in 10 dimensions and the
dualities that  relate them to M theory in 11 dimensions. 
Then I will mention some important unresolved issues.
\end{abstract}

\section*{Introduction}
Superstring theory first achieved widespread acceptance during the {\em first
superstring revolution} in 1984-85. There were three main developments at
this time.  The first was the discovery of an anomaly cancellation mechanism \cite{green84},
which showed that supersymmetric gauge theories can be consistent in ten dimensions
provided they are coupled to supergravity (as in type I superstring theory)
and the gauge group is either SO(32) or $E_8 \times E_8$. Any other group
necessarily would give uncanceled gauge anomalies and hence inconsistency at
the quantum level.  The second development was the discovery of two new
superstring theories---called {\em heterotic} string theories---with precisely these
gauge groups \cite{gross84}.  The third development was the realization that the $E_8 \times
E_8$ heterotic string theory admits solutions in which six of the space dimensions
form a Calabi--Yau space, and that this results in a 4d effective theory at low
energies with many qualitatively realistic features \cite{candelas85}.  
Unfortunately, there are
very many Calabi--Yau spaces and a whole range of additional choices
that can be made (orbifolds,
Wilson loops, etc.).  Thus there is an enormous variety of possibilities, none of
which stands out as particularly special.

In any case, after the first superstring revolution subsided, we had five
distinct superstring theories with consistent weak coupling perturbation
expansions, each in ten dimensions.  Three of them, the {\em type I} theory and
the two heterotic theories, have ${\mathcal N} = 1$ supersymmetry in the ten-dimensional
sense.  Since the minimal 10d spinor is simultaneously Majorana and Weyl, this
corresponds to 16 conserved supercharges.  The other two theories, called {\em type
IIA} and {\em type IIB}, have ${\mathcal N} = 2$ supersymmetry (32 supercharges)~\cite{green82}.  
In the IIA case the two spinors have opposite handedness so that the spectrum is
left-right symmetric (nonchiral).  In the IIB case the two spinors have the
same handedness and the spectrum is chiral.

The understanding of these five superstring theories was developed in the
ensuing years.  In each case it became clear, and was largely proved, that
there are consistent perturbation expansions of on-shell scattering amplitudes.
In four of the five cases (heterotic and type II) the fundamental strings are
oriented and unbreakable.  As a result, these theories have particularly simple
perturbation expansions.  Specifically, there is a unique Feynman diagram at
each order of the loop expansion.  The Feynman diagrams depict string world sheets, and
therefore they are two-dimensional surfaces.  For these four theories the
unique $L$-loop diagram is a closed orientable
genus-$L$ Riemann surface, which can be visualized
as a sphere with $L$ handles.  External (incoming or outgoing) particles are
represented by $N$ points (or ``punctures'') on the Riemann surface.  A
given diagram represents a well-defined integral of dimension $6L + 2N - 6$.  This
integral has no ultraviolet divergences, even though the spectrum contains
states of arbitrarily high spin (including a massless graviton).  From the
viewpoint of point-particle contributions, string and supersymmetry properties
are responsible for incredible cancellations.  Type I superstrings are
unoriented and breakable.  As a result, the perturbation expansion is more
complicated for this theory, and the various world-sheet diagrams at a given order 
(determined by the Euler number) have to be
combined properly to cancel divergences and anomalies ~\cite{green85}.

\section*{M Theory}

In the 1970s and 1980s various supersymmetry and 
supergravity theories were constructed. (See~\cite{salam}, for example.)
In particular, supersymmetry representation theory showed that ten is the largest
spacetime dimension in which there can be a supersymmetric  Yang--Mills
theory, with spins $\leq 1$ \cite{brink77}.  
This is a pretty ({\it i.e.}, very symmetrical) classical field theory, but
at the quantum level it is both nonrenormalizable and 
anomalous for any nonabelian gauge group.  However, as
we indicated earlier, both problems can be overcome for suitable gauge groups
(SO(32) or $E_8 \times E_8$) when the Yang--Mills theory 
is embedded in a type I or heterotic string theory.

The largest possible spacetime dimension for a supergravity theory (with spins $\leq 2$),
on the other hand, is eleven.  Eleven-dimensional supergravity, which has 32 conserved
supercharges, was constructed 20 years ago \cite{cremmer78a}.  
It has three kinds of fields---the
graviton field (with 44 polarizations), the gravitino field (with 128
polarizations), and a three-index 
antisymmetric tensor gauge field $C_{\mu\nu\rho}$ (with 84
polarizations).  These massless particles are referred to
collectively as the {\em supergraviton}. 
11d supergravity is also a pretty classical field theory, which has attracted
a lot of attention over the years.  It is not chiral, and therefore not subject
to anomaly problems.\footnote{Unless the spacetime has boundaries.  The
anomaly associated to a 10d boundary can be canceled by introducing $E_8$
supersymmetric gauge theory on the boundary \cite{horava95}.}  
It is also nonrenormalizable, and thus it cannot be a fundamental theory. 
However, we now
believe that it is a low-energy effective description of M theory, which is a
well-defined quantum theory~\cite{witten95a}.  
This means, in particular, that higher dimension
terms in the effective action for the supergravity fields have uniquely determined
coefficients within the M theory setting, even though they are formally 
infinite 
(and hence undetermined) within the supergravity context.

Intriguing connections between type IIA string theory and 11d supergravity have
been known for a long time.  If one 
carries out {\em dimensional reduction} of 11d supergravity
to 10d, one gets type IIA supergravity~\cite{campbell84}.  
In this case dimensional reduction can be viewed as
a compactification on a circle in which one drops all the Kaluza--Klein
excitations.  It is easy to show that this does not break any of the
supersymmetries. The field equations of 11d supergravity admit a solution that describes a
supermembrane.  This solution has the property that the energy
density is concentrated on a two-dimensional surface.  A 3d world-volume
description of the dynamics of
this supermembrane, quite analogous to the 2d world volume
actions of superstrings, has been constructed~\cite{bergshoeff87}.  
The authors suggested that a
consistent 11d quantum theory might be defined in terms of this membrane, in analogy
to string theories in ten dimensions.\footnote{It is now clear that this cannot be done 
in any straightforward manner, since there is no weak coupling limit in which
the supermembrane describes all the finite-mass excitations.}  
Another striking result was the discovery of
double dimensional reduction~\cite{duff87}.  
This is a dimensional reduction on a circle, in which one
wraps one dimension of the membrane around the circle
and drops all Kaluza--Klein excitations for both the spacetime theory and the
world-volume theory.  The remarkable fact is that this gives the (previously
known) type IIA superstring world-volume action~\cite{green84b}.

For many years these facts remained unexplained curiosities until they were
reconsidered by Townsend~\cite{townsend95a} and by Witten~\cite{witten95a}.  
The conclusion is that type IIA
superstring theory really does have a circular 11th dimension in addition to
the previously
known ten spacetime dimensions.  This fact was not recognized earlier
because the appearance of the 11th dimension is a nonperturbative phenomenon,
not visible in perturbation theory.

To explain the relation between M theory and type IIA string theory, a good
approach is to identify the parameters that characterize each of them and to
explain how they are related.  Eleven-dimensional supergravity (and hence M
theory, too) has no dimensionless parameters.  As we have seen, there are no
massless scalar fields, whose vevs could give parameters.  The only
parameter  is the 11d Newton constant, which raised to a suitable power 
($-1/9$), gives the 11d Planck mass $m_p$.  
When M theory is compactified on a
circle (so that the spacetime geometry is $R^{10} \times S^1$) another
parameter is the radius $R$ of the circle.
The parameters of type IIA superstring theory are the
string mass scale $m_s$, introduced earlier, and the dimensionless string
coupling constant $g_s$.  An important fact about all five superstring theories
is that the coupling constant is not an arbitrary parameter.  Rather, it is a
dynamically determined vev of a scalar field, the {\em dilaton,} which is a
supersymmetry partner of the graviton.  With the usual conventions, one has
$g_s = \langle e^\phi\rangle$.

We can identify compactified M theory with type IIA superstring theory by
making the following correspondences:
\begin{equation}\label{M1}
m_s^2 = 2\pi R m_p^3
\end{equation}
\begin{equation}\label{M2}
g_s = 2\pi Rm_s.
\end{equation}
Conventional string perturbation theory is an expansion in powers of $g_s$ at
fixed $m_s$.  Equation~(\ref{M2}) shows that this is equivalent to an expansion
about $R=0$.  In particular, the strong coupling limit of type IIA superstring
theory corresponds to decompactification of the eleventh dimension, so in a sense M theory
is type IIA string theory at infinite coupling.\footnote{The $E_8 \times E_8$ heterotic
string theory is also eleven-dimensional at strong coupling \cite{horava95}.}
This explains why the eleventh dimension was not discovered in
studies of string perturbation theory.

These relations encode some interesting facts.  The fact relevant to eq.~(\ref{M1})
concerns the interpretation of the fundamental type IIA string.
Earlier we discussed the old notion of double dimensional reduction, which
allowed one to derive the IIA superstring world-sheet action from the 
11d supermembrane (or M2-brane)
world-volume action.  Now we can make a stronger statement:  The fundamental
IIA string actually {\em is} an M2-brane of M theory with one of its dimensions
wrapped around the circular spatial dimension.   No truncation to zero modes is
required. Denoting the string and membrane tensions (energy
per unit volume) by $T_{F1}$ and $T_{M2}$, one deduces that
\begin{equation}
T_{F1} = 2\pi R \, T_{M2}.
\end{equation}
However, $T_{F1} = 2\pi m_s^2$ and $T_{M2} = 2\pi m_p^3$.  Combining these
relations gives eq.~(\ref{M1}). It should be emphasized that all the formulas in this
section are exact, due to the large amount of unbroken supersymmetry.

Type II superstring theories contain a variety of $p$-brane solutions that preserve
half of the 32 supersymmetries. These are solutions in which
the energy  is concentrated on a $p$-dimensional spatial
hypersurface. (Adding the time dimension, the  world volume 
of a $p$-brane has $p+1$ dimensions.)
The corresponding solutions of  supergravity
theories were constructed by Horowitz and Strominger~\cite{horowitz91}.
A large class of these $p$-brane excitations are called
{\em D-branes} (or D$p$-branes when we want to specify the dimension),
 whose tensions are given by~\cite{polchinski95}
\begin{equation} \label{Dtension}
T_{Dp} = 2\pi {m_s^{p+1}}/{g_s}.
\end{equation}
This dependence on the coupling constant is one of the characteristic features
of a D-brane.  It is to be contrasted with the more familiar 
$g^{-2}$ dependence of soliton masses
(e.g., the 't Hooft--Polyakov monopole).
Another characteristic feature of D-branes
is that they carry a charge that couples to a gauge field
in the Ramond-Ramond (RR) sector of the theory. 
(Such fields can be described as bispinors.)
The particular RR gauge fields that 
occur imply that
even values of $p$ occur in the IIA theory and odd values in the IIB theory.

D-branes have a number of special properties, which make them especially
interesting.   By definition, they are branes on which strings can end---D
stands for {\em Dirichlet} boundary conditions.  The end of a string carries a
charge, and the D-brane world-volume theory contains a $U(1)$ gauge field that carries the
associated flux.  When $n$ D$p$-branes are coincident, or parallel and nearly
coincident, the associated $(p + 1)$-dimensional world-volume theory is a
$U(n)$ gauge theory.  The $n^2$ gauge bosons $A_\mu^{ij}$ and their
supersymmetry partners arise as the ground states of oriented strings running
from the $i$th D$p$-brane to the $j$th D$p$-brane.  The diagonal elements,
belonging to the Cartan subalgebra, are massless.  The field
$A_\mu^{ij}$ with $i \not= j$ has a mass proportional to the separation of the
$i$th and $j$th branes.  This separation is described by the vev of a
corresponding scalar field in the world-volume theory.

In particular, the D2-brane of the type IIA theory
corresponds to our friend the supermembrane of M theory, but now
in a background geometry in which one of the transverse dimensions is a circle.
The tensions check, because (using eqs.~(\ref{M1}), (\ref{M2}), and (\ref{Dtension}))
$T_{D2} = 2\pi {m_s^3}/{g_s} = 2\pi m_p^3 = T_{M2}$.
The mass of the first Kaluza--Klein excitation 
of the 11d supergraviton is $1/R$.  Using eq.~(\ref{M2}),
we see that this can be identified with the D0-brane.
More identifications of this type arise when we consider the magnetic dual of
the M theory supermembrane.  This turns out to be a five-brane, called the
M5-brane.\footnote{In general, the magnetic dual of a $p$-brane in $d$
dimensions is a $(d - p - 4)$-brane.}  Its tension is $T_{M5} = 2\pi m_p^6$.
Wrapping one of its dimensions around the circle gives the D4-brane, with
tension $T_{D4} = 2\pi R \,T_{M5} = 2\pi m_s^5/g_s$.
If, on the other hand, the M5-frame is not wrapped around the circle, one
obtains the so-called NS5-brane of the IIA theory with tension
\[
T_{NS5} = T_{M5} = 2\pi m_s^6/g_s^2.
\]
This 5-brane, which is the magnetic dual of the fundamental IIA string, exhibits
the conventional $g^{-2}$ solitonic dependence.

To summarize, type IIA superstring theory is M theory compactified on a circle
of radius $R=g_s \ell_s$.
 M theory is believed to be a well-defined quantum theory in 11d, which is
approximated at low energy by 11d supergravity.  Its 
supersymmetric excitations (which are the only ones known when there is no
compactification) are the
massless supergraviton, the M2-brane, and the M5-brane.  These account both for
the (perturbative) fundamental string of the IIA theory and for many of its
nonperturbative excitations.  The identities presented here are exact,
because they are protected by supersymmetry.

\section*{Important Unresolved Issues}

One issue that needs to be settled if superstring theory is to be used for
phenomenology is where supersymmetry fits into the story. It is
clear that at the string scale ($\approx 10^{18}$ GeV) or the Planck scale the underlying theory has  maximal supersymmetry (32 conserved supercharges). The question that needs to
be answered is at what scales they are broken and by what mechanisms.
The traditional picture (which looks the most plausible to me) is that 
at the compactification/GUT scale ($\approx 10^{16}$ GeV) the symmetry is
broken to ${\mathcal N} = 1$ in $d =4$ (four conserved supercharges), and this
persists to the TeV scale, where the final susy breaking occurs. The TeV scale is
indicated by three separate arguments: the gauge hierarchy problem, supersymmetric
grand unification, and the requirement that the lightest superparticle (LSP) be
a cosmologically significant component of dark matter. It would be astonishing
if this coincidence turned out to be a fluke. There is other support for this picture
such as the mass of the top quark and the ease with which it
gives electroweak symmetry breaking. Despite all these
indications, we cannot be certain that this picture is correct until it is demonstrated
experimentally. As I once told a newspaper reporter: discovery of
supersymmetry would be more profound than life on Mars. 

Another important issue is the problem of vacuum degeneracy and the
stabilization of moduli. Let me explain. The underlying theory is completely
unique, with no dimensionless parameters. Nevertheless, typical quantum vacua
have continuous parameters, called {\em moduli}, which arise as the vacuum
values of scalar fields. Notable examples are the sizes of extra dimensions and the
string coupling constant. For typical string vacua, the effective potential has many
flat directions, so there is a continuum, or {\em moduli space}, of minima. The
fields that correspond to the flat directions describe massless spin zero particles.
These particles typically interact with roughly gravitational strength, which is a 
problem because the gravitational force is observed to be pure tensor to better
than 1\% accuracy. So it seems that we should seek a vacuum without moduli,
which is very difficult to do. However, if a realistic vacuum of this type is ever found, 
it will not have any continuously adjustable parameters, and therefore it will
be completely predictive (at least in principle).

Perhaps the most challenging unresolved issue of all, is the {\em cosmological constant}.
This is a term in the effective action that describes the energy
density of the vacuum, which is observable in a gravitational theory. 
Observationally, there are indications that
it may be nonzero, but it is extremely small. Taking the
$1/4$ power, the energy scale is $\leq 10^{-11}$ GeV. In a fundamental theory
it receives contributions from many sources such as vacuum condensates and
zero point energies. Supersymmetry ensures that boson and fermion zero point
energies cancel, so the natural scale would seem to be the TeV susy breaking scale,
which  is many orders of magnitude too high. This is a fine-tuning problem
that is reminiscent of the gauge  hierarchy problem. Presumably string theory
will provide an elegant solution. Until we know what the relevant mechanism
is, it is hard to be confident that there is not an alternative to supersymmetry
for solving the gauge hierarchy problem. I believe that when the correct solution
to the problem of the cosmological constant is found, it will spark another
revolution in our understanding. Recently toy models without supersymmetry
that seem to have a vanishing cosmological constant have been constructed by
Kachru, Kumar, and Silverstein~\cite{kachru}.  These models are far from realistic,
and have not led yet to new qualitative understandings. However, they are 
the best lead we have at the present time.

To conclude, there has been dramatic progress in understanding string theory in the
past few years, but there are still crucial issues that remain unresolved. Future
experimental discoveries will be essential to help guide our thinking. Sooner (Tevatron)
or later (LHC) exciting new phenomena are bound to show up. My bet is on Higgs and
superparticles. But if I should turn out to be wrong, 
that would not mean that string theory is wrong.

\end{document}